# Ultrafast spectroscopy of quasiparticle dynamics in cuprate superconductors

Wei Li,[1] Chunfeng Zhang,[1,*] Xiaoyong Wang,[1] Jak Chakhalian,[2,*] and Min Xiao[1,2,*]

[1]National Laboratory of Solid State Microstructures and Department of Physics, Nanjing University, Nanjing 210093, China
[2]Department of Physics, University of Arkansas, Fayetteville, Arkansas 72701, USA

E-mail: cfzhang@nju.edu.cn, jchakhal@uark.edu, mxiao@uark.edu

**Abstract**

Ultrafast pump-probe spectroscopy is a powerful tool to study the nonequilibrium dynamics in high-Tc cuprate superconductors. The photo-induced quasiparticle (QP) dynamics revealed by pump-probe spectroscopy are sensitive to the near-Fermi level electronic structures. Here we review several selected examples to illustrate the enduring challenges including pairing glue, phase separation, and phase transitions in cuprate superconductors. We also present the data obtained on thin films of $YBa_2Cu_3O_{7-\delta}$ in connection to these issues.

## 1. Introduction

The nature of high-temperature superconductivity (HTSC) in cuprate compounds has been a central challenge in modern condensed-matter physics [1]. The past two decades of vigorous research has uncovered a plethora of exciting phenomena mapped on to the complex and surprisingly asymmetric (electron vs. hole doping) phase diagrams. To understand the microscopic mechanisms, it is essential to elucidate the physics behind the exotic properties and at the same time to identify those features that are quintessential to the phenomena of HTSC. Despite great successes in unraveling the electronic and magnetic properties of cuprates by time-integrated techniques, such as electric and thermal transport [2], tunneling [3], angle-resolved photoemission (ARPES) [4], and conventional optical spectroscopies [5], there are still unresolved challenges in disentangling the interplay between the elementary excitations. To this end, ultrafast optical spectroscopy has been extensively exploited as a unique probe with the potential to explore the physics of elementary excitations on a temporal scale by elucidating spectrotemporal features for specific physical processes [6-8].

Historically, since the early 80s, the technique of pump-probe femtosecond spectroscopy has been applied as a tool for investigating nonequilibrium dynamics in conventional metals and semiconductors [9, 10]. As an example, it was able to provide the earliest reliable experimental evaluation of the electron-phonon renormalization parameter [9]. This extraordinary achievement stimulated rapid interest in using the femtosecond laser as a probe of high temperature superconductivity [11-17] and more generally correlated physics in complex oxides materials [6, 18, 19]. Towards this end, in the early 1990s, some pioneering works began on high-Tc cuprate superconductors [11-17]. In the past three decades, pump-probe spectroscopy has been successfully applied to study the QP dynamics in many cuprate superconductors [20-22].

From the experimental point of view, the technique relies on a fractional change in reflectivity or transmission monitored as a function of pump-probe delay time. The photo-induced reflectivity change ($\Delta R/R$) is assumed to be proportional to the density of photo-excited quasiparticles (QPs) ( $\Delta N(t)$ ) [12, 23], i.e., $\Delta R/R(t) \propto \Delta N(t)$, since the change in dielectric constant ($\Delta \varepsilon$) is linearly dependent on the QP density as $\Delta \varepsilon = (1 + i/\omega \tau_s) \omega_p^2 \Delta N_0 / N_t \omega^3$, where $\omega$ is the frequency of probe photon, $\omega_p$ is the plasma frequency, $\tau_s$ is the scattering time, and $N_t$ is the total electron density [12, 24]. When temperature surpasses the superconducting transition temperature ($T_C$), the magnitude of $\Delta R/R$ exhibits an abrupt drop [25]. The recovery traces can be analyzed as multiple exponential decay components with each decay channel assigned to a specific physical process [8, 26-28].

Recently, thanks to the rapid advances in laser technology, significant progresses have been achieved in ultrafast spectroscopic studies of superconductors [8, 27, 29-34]. Here, we want to specially mention the development of Ti:Sapphire lasers which has enabled the pump-probe method at 800/400 nm and resulted in fruitful results on superconductors [35-39]. At the same time, the optical response at 800 nm can only give indirect information due to the probe photon energy being much larger than the energy scale specific to order parameters and elementary excitations in superconductors [5, 24]. To address this issue, broadband pump-probe spectroscopy based on the super-continuum generation and optical parametric amplification has been employed [8, 27, 40, 41]. This approach is able to provide dynamical information in both time and frequency domains, connecting the reflectivity change at high-energy electronic excitation with the resonant spectral function of Boson modes [8]. Furthermore, it is now feasible to detune ultrashort pulses to the extreme wavelength regime. For example, with different nonlinear optical methods on a lab bench, the spectral coverage from deep ultraviolet to terahertz (THz) bands is now available, leading to the rapid developments of time-resolved infrared (IR) - THz spectroscopy[6, 42], time-resolved ARPES [30, 31], and time-resolved Raman spectroscopy [43]. These new probes have dramatically improved the capabilities of the ultrafast optical techniques, shifting interests to resonant excitations and probes. To illustrate, the technique of high harmonic generation provides deep UV pulses that have been integrated into ARPES; such time-resolved ARPES enables mapping of the transient density of states dynamically in cuprates, which may answer some enduring issues in such interesting systems [30, 31].

Unlike the rapid development in ultrafast experimentation on superconductors, to date the progress in microscopic theory has been somewhat limited. Early on, the non-equilibrium phenomena in conventional superconductors were described by Rothwarf and Taylor in 1967 [44]. Such a phenomenological model, expressed in the set of the Rothwarf-Taylor equations (RTE), has been successfully applied for BCS superconductors and is frequently employed to interpret QP dynamics in cuprates. Later on, Kabanov and co-workers have suggested to treat the ground state with two types of gaps: the temperature-dependent gap (superconducting gap) and the temperature-independent gap (pseudogap) [25]. Their model has successfully explained the temperature dependence of signal amplitude and decay lifetime. Nevertheless, the dynamic behavior in cuprates was found to be more consistent with the model of an isotropic gap [25], which is in contrast to the general argument of a d-wave symmetry gap in cuprates. Nicol and Carbotte attempted to model the QP dynamics in cuprates with two nonequilibrium models (the $\mu^*$ and $T^*$ models) [45], but they found that neither s-wave nor d-wave gap models can produce a quantitative explanation to the observed behaviors. Recently, there have been new proposals to describe the nonequilibrium phenomena in high temperature superconductors [46-48]. For experimental scientists, phenomenological descriptions, such as two/three/four-temperature models, in the view of thermodynamics have been widely used to fit the experimental data [8, 10, 49].

The QP dynamic behaviors revealed by pump-probe experiments in cuprates are tightly associated with the spectral weight transfer due to the photo-induced modification of ground state. The recovery dynamics have been studied to investigate glue for Cooper pairing including phononic and electronic contributions. Initially, ultrafast spectroscopy has been performed on $YBa_2Cu_3O_{7-\delta}$ (YBCO) [11-15, 21, 50, 51]; ever since, the study has been expanded to include a variety of compounds in the cuprate family [8, 17, 20, 27, 29, 39, 40, 49, 52-55]. The normal state has also been at the center of focus to investigate the QP dynamics related to the pseudogap phase and other competing orders [34, 37, 38]. The ultrafast optical spectroscopy is expected to provide important information on this long-standing issue by searching for distinct spectrotemporal features of superconductivity and the pseudogap. These features have been captured in reflection change, spectral dispersion, polarization dependence and coherent phonon anomalies [20, 25, 40, 56, 57].

In addition to probing the QPs at different phases in cuprates, light-driven phase transition has also attracted a rapidly-growing interest [18]; e.g. photo-induced suppression and enhancement of superconductivity has been recently explored [23, 58]. In this paper, we briefly highlight the achievements in the ultrafast optical studies of cuprates with an emphasis on the QP dynamics related to phase transitions. Special emphasis will be placed on selected results associated with pairing glues, phase separation, light-induced phase transitions, and the present data we have recently obtained on thin films of YBCO in connection to these issues.

## 2. Experiments

To date, the majority of time-resolved studies on cuprates were based on the pump-probe configuration schematically shown in Fig. 1(a). In this configuration, the output of a femtosecond laser is split into a pump and a probe beams with the probe beam always set to be much weaker than the pump one. After pre-designed wavelength detuning, the two beams are focused onto a same sample spot. Since HTSC is a low T phenomenon, a sample is usually mounted inside an optical cryostat. The change in reflection or transmission of the probe beam is recorded as a function of the time delay between the pump and probe pulses. Previously, femtosecond pulses with different repetition rates have been used. In general, a higher repetition rate provided by an oscillator may yield a better sensitivity ($\Delta R/R \sim 10^{-7}$) that is capable of probing the QP dynamics in a weak perturbation regime. The steady laser heating, however, may cause undesired artifacts to appear in spectra so it is usually necessary to reduce the repetition rate with a pulse selector. For experiments with low repetition rates, the pulses from a regenerative amplifier can be employed. The pulse energy can be orders of magnitude higher at the expense of sensitivity (e.g. $\Delta R/R \sim 10^{-5}$ for 1 kHz). With high energy pulses, it is feasible to detune the pulses with a broadband wavelength coverage through a variety of nonlinear optical techniques such as super-continuum generation, optical parametric amplification, high harmonic generation and terahertz generation. These technical advances enable resonant excitations of different modes. As a result, on the probe side, broadband visible-NIR reflectivity [8, 27], terahertz optical conductivity [6, 59], circular dichroism [60], and ARPES [31] are already available for experimentation.

To illustrate some of the experimental challenges, we present our results obtained with the pump-probe spectroscopy on a nearly optimally doped YBCO film with a thickness of 48 unit cells. The samples are grown on an atomically flat single crystal $SrTiO_3$ (001) substrate in layer-by-layer fashion by laser MBE [61, 62]. As characterized by the transport measurement, the sample exhibits the superconductivity transition at 93 K and the pseudogap phase around 152 K where the resistance deviates from the linear T regime [62]. The pump-probe measurements were conducted using a regenerative amplifier (Libra, Coherent Inc) with a repetition rate set at 1 kHz. For measurement with broadband spectral coverage, the probe wavelength was detuned by an optical parametric amplifier (OperA Solo, Coherent Inc); we also employed the technique of balanced detection together with a lock-in amplifier to improve the sensitivity.

**3. Phenomena in Cuprates**

Pump-probe traces recorded from cuprate superconductors typically exhibit a photo-excited abrupt change in reflectivity followed by a recovery decay consisting of one or multiple exponential components. The amplitude and lifetime parameters of these components are linked to the electronic and/or phononic responses. These features have been studied as functions of excitation power, probe wavelength, temperature, an applied magnetic field and doping level to address key problems of high-Tc superconductivity. To this end, the natures of pairing glue, competing orders, phase transitions and other critical issues have been surveyed [6, 8, 12, 14, 16, 17,

20-23, 25-31, 37, 38, 40, 42, 43, 45, 48-50, 52, 53, 55-59, 63-71]. In addition to the exponential time decays, in some cases, coherent oscillations have been reported; the oscillation components have been assigned to different collective modes such as coherent phonons [37, 41, 63], charge-density waves (CDWs) [34], and Josephson plasma solitons [32].

The exact nature of photo-excitations in cuprates is still under intensive study. At the present time, a generally accepted picture involves the process of avalanche multiplication where an absorbed photon generates an electron-hole pair (Fig. 1b). The electron-hole pair then loses its energy by exciting multiple QPs via the process of avalanche multiplication on the ultrafast time scale. In the superconducting phase, the number of QPs created per absorbed photon can be evaluated approximately by the ratio between photon energy ($E_{ph}$) and the superconducting gap ($\Delta$). In addition, the temperature dependences of signal magnitude have been explored by varying excitation power in weak and intense regimes, respectively. In the weak excitation regime, the density of photo-injected QPs is substantially lower than the carrier density at the normal state. Based on this the amplitude of the photo-induced signal is assumed to be linearly dependent on the density of QPs. In this regime, the QP dynamics under excitation is believed to reflect intrinsic electronic behaviors of the cuprates. Upon increasing the excitation power, high density photons start to destroy the superconducting condensate and a photo-induced phase transition may take place with saturation of the signal amplitude. In addition to the signal amplitude, the decay lifetime of photo-excited QPs is strongly affected by the interaction between QPs and elementary excitations [25, 44]. That explains why the recovery dynamics can provide important information about microscopic interactions in superconductors.

### 3.1. Pairing glue

Despite the fundamental difference between cuprates and conventional BCS superconductors, it is widely believed that the Cooper-pair formation is of equivalent importance in both of these systems. One of the key challenges here is to identify the nature of the glue (or mediator) that is responsible for Cooper-pair formation [4, 24, 72]. Lattice vibrations (phonons) and electronic boson modes have been considered as potential glues for pairing in cuprates. Presently, the pure phonon-mediated pairing has been rarely considered as the major glue mechanism because of its failure to account for many observed phenomena in cuprates including the weak isotope effect [54, 73], d-wave gap symmetry [2], and high transition temperature. Nevertheless, convincing evidences for significant electron-phonon (e-ph) contributions have been alluded to integrated experiments with ARPES [73-75], inelastic neutron scattering [76], tunneling [77] and Raman [78] spectroscopy. For example, electron-boson coupling has been associated with an ubiquitous kink at ~ 70 meV in the QP dispersion curve as measured by ARPES [4]. Unfortunately, there is certain ambiguity in assigning it to the phononic or electronic degree of freedom by the time-integrated techniques since the resonant energies of optical Cu-O lattice modes [74] and spin excitation modes [79] overlap.

As a complementary route, to disentangle these interactions on a temporal scale, ultrafast spectroscopy has been conducted to address this fundamental issue [8, 14, 28, 64]. To understand the roles played by these interactions, it is essential to evaluate their coupling strengths. The interaction between the fermionic QPs with bosonic excitations, i.e. phonons and spin fluctuations, has been described by the bosonic function $\Pi(\Omega)$ [ $\alpha^2 F(\Omega)$ for e-ph coupling and $I^2\chi(\Omega)$ for spin fluctuations] within the Migdal-Eliashberg theory [80]. The strength of this e-ph coupling is then quantified in terms of $\lambda<\Omega^2>=2\int \alpha^2 F(\Omega)\Omega d\Omega$; such defined e-ph coupling parameter, $\lambda$, plays a central role in determining the transition temperature $T_C$.

Among first attempts to evaluate the value of $\lambda<\Omega^2>$ with pump-probe experiments, Allen [5] considered a phenomenological two-temperature model. In this model the relaxation lifetime ($\tau_{e-ph}$) of QPs is connected to the e-ph coupling strength as $\lambda<\Omega^2>=\pi k_B T_e / 3\hbar \tau_{e-ph}$, where $T_e$ is the electronic temperature that is much higher than the lattice temperature $T_L$. In addition, the lifetime parameter for electron-electron (*e-e*) interaction is assumed to be much faster than that of the *e-ph* interaction, i.e. $\tau_{e-e} \ll \tau_{e-ph}$. Based on this model, an analysis of experimental data on YBCO allowed the estimation of the parameter in terms of $\lambda<\Omega^2> \sim 400$ meV with the e-ph coupling parameter $\lambda \sim 0.9$ [14]. As a further development, Perfetti *et al.* extended the two-temperature model to derive the strength of *e-ph* interaction from the time-resolved ARPES data recorded from the sample of $Bi_2Sr_2CaCu_2O_{8+\delta}$ [49]; from their analysis the phonons were divided into the strongly-coupled modes and the weakly-coupled modes. Within this three-temperature model, they concluded that only twenty percents of phonons are strongly coupled to the QPs. The energy exchange between the QPs and strongly-coupled phonon modes occurs in a time scale of ~ 110 fs. The value of $\lambda<\Omega^2>$ was estimated to be ~ 380 meV$^2$ with $\lambda \sim 0.25$.

On the other hand, later development of the theory of *e-ph* interaction in metals resulted in an analytical solution of the non-equilibrium model [81, 82] and yielded a relation of $\lambda<\Omega^2> = 2\pi k_B T_L / 3\hbar \tau_{e-ph}$, which predicts that the strength of *e-ph* coupling depends on the lattice temperature $T_L$ rather than the electron temperature. Gadermaier *et al.* conducted pump-probe experiments on samples of $YBa_2Cu_3O_{6.5}$ and $La_{1.85}Sr_{0.15}CuO_4$ with an improved temporal resolution of ~ 20 fs [28]. They concluded that the non-equilibrium model provides a better description of the *e-ph* interaction since the observed QP dynamics was found to be independent of the excitation density. The spectral function $\lambda<\Omega^2>$ was estimated to be ~ 400 meV$^2$ and ~ 800 meV$^2$ in $YBa_2Cu_3O_{6.5}$ and $La_{1.85}Sr_{0.15}CuO_4$, respectively. The results argued for the possibility of a polaronic pairing mechanism. Additionally, Pashkin *et al.* carried out a transient optical study on an optimally doped YBCO crystal with resonantly probing the phonons by utilizing a broadband THz source [69]. They observed that hot phonons were excited on the same temporal scale of ~ 150 fs as the QPs, implying that the *e-ph* interaction may be much stronger than usually assumed.

In parallel to the works on the *e-ph* interaction, other groups have performed transient optical experiments to explore electronic pairing glue. From a pump-probe experiment in mid-infrared, Kaindl *et al.* found that the electronic excitations were correlated with the 41 meV antiferromagnetic spin fluctuations having the same temperature dependent behavior in YBCO superconductors [64]. Very recently, Dal Conte *et al.* performed a broadband pump-probe study at room temperature focusing on the dynamics of electron-boson coupling in $Bi_2Sr_2Ca_{0.92}Y_{0.08}Cu_2O_{8+\delta}$ [8]. From the spectral dispersion of $\Delta R/R$, they argued that there is an electron-boson interaction that is much stronger than the interaction between the electron and strongly-coupled phonons. This strong interaction was assigned to the glue contributed by electronic correlations, e.g. spin fluctuations or current loops. In addition, they introduced a four-temperature model by adding another energy dissipation process to the QPs due to the strong coupling between QPs and electronic boson modes. By further extending this four-temperature model, Chia *et al.* introduced a five-temperature model in which the QPs were assumed to be coupled to four boson modes, i.e. the strongly/weakly-coupled phonon/spin fluctuations. After careful evaluation of the strength of each coupling, the authors observed strong doping dependences of *e-ph* and electron-spin fluctuation couplings in the $Bi_2Sr_2CaCu_2O_{8+\delta}$ system [83].

Despite the remarkable progress made in recent years, to date the debate on whether the major glue is phononic or electronic seems to remain unsettled. In principle, time-resolved spectroscopy can play an important role in tackling this challenge since it can reliably separate different interactions on temporal scale [8] that is unattainable for most other complementary probes. We also note that, until now, most of the transient optical measurements have been performed in the wavelength domain far from the energy scale of boson modes and with the temporal resolving power limited by the pulse duration of modern lasers. One can envision, however, that with the rapidly-developing technique for few-cycle pulses [84], along with the infrared resonant probes [6, 69] and time domain ARPES [31], it may become possible to accurately monitor the coupling strengths between electrons and each boson modes.

**3.2 Phase separation**

It has been established early on that upon doping cuprate superconductors exhibit a physics rich phase diagram [85]. At low temperatures, under-doped cuprates show an antiferromagnetic phase and superconductivity at intermediate doping level. Some other orderings, such as spin-glass, charge-density waves (CDWs) and pseudogap phase, may also appear in the proximity to superconductivity. For the normal state at temperatures above $T_C$, the pseudogap phase emerges and persists down to a characteristic temperature $T^*$. The intriguing relationship between the pseudogap and superconductivity is one of the most discussed issues in the physics of cuprate superconductors.

Ultrafast laser spectroscopy has been employed to investigate each phase in cuprates. For the YBCO family, it was found that dynamics of photo-excited QPs are strongly dependent on the level of doping [20-22, 67]. Specifically, in the near optimally doped sample, the magnitude of signal shows an abrupt change at $T_C$ when the excitation power is weak. However, it was found that the signal $\Delta R/R$ persists to higher temperature in under- and over-doped samples due to the presence of the pseudogap. To this end, Kabanov *et al.* considered the recombination process with the scenario of the gap formation close to the Fermi surface and obtained a quantitative explanation [25]. With a temperature-dependent superconducting gap $\Delta(T)$, they found good agreement with experimental data for optimally doped YBCO assuming an isotropic gap. The analysis was done according to the following equation

$$\frac{\Delta R}{R} \propto \frac{1}{(\Delta(T)+k_B T/2)\{1+g\sqrt{k_B T/\Delta(T)}\exp[-\Delta(T)/k_B T]\}}, \quad (1)$$

where $g$ represents the ratio of bosonic and electronic density of states that contribute to the QP density, and $k_B$ is the Boltzmann constant. Interestingly, this model was also successfully used in experiments on iron-based superconductors [68, 86]. Despite reasonable fit to experimental data, the model relies on the assumption of s-wave gap symmetry that is in conflict with the general belief about the d-wave gap symmetry in cuprates. To resolve this discrepancy, Segre *et al.* proposed another explanation of the pump-probe results in under-doped cuprates [21]. Considering the d-wave gap symmetry, they suggested that besides recombination via the gap at the anti-nodal region, the photo-excited QPs might also relax through the thermalization process via nodal points.

In addition to the amplitude, the recovery dynamics also changes abruptly near $T_C$. Early on, it was observed that the relaxation becomes much slower near $T_C$, which was explained as a result of the phonon bottleneck effect with the extended RTE model [44]. Such anomalies of QP relaxation have also been found in several other systems [86]. Within the RTE model, the magnitude of the superconducting gap can be connected with $\Delta R/R$ [27]. To investigate the gap symmetry, Luo *et al.* conducted a polarization-dependent ultrafast study on YBCO crystals [87]. They found that the QP relaxation is polarization dependent and the bottleneck relaxation process is strongly anisotropic in superconducting states when interpreted as a signature of d-wave gap symmetry.

From the experiments on under- or over-doped samples, it was concluded that there is another faster component whose presence is linked to the temperature-independent pseudogap. The pseudogap feature persists to a temperature $T^*$ that varies with the level of doping in cuprates. Since a pump-probe measurement is sensitive to the gap dynamics near the Fermi surface, it can be used to characterize the pseudogap. The measurements in this direction done by Mihailovic's group have found that the $\Delta R/R$ signal in YBCO samples with different doping levels can be well reproduced by a sum of responses from the superconducting gap and pseudogap [20, 25]. The spectral weight transfer induced by photo excitation also shows a strong dependence on the level of doping; namely the transient traces of photo-induced $\Delta R/R$ exhibit a sign change in moving from the under-doped to over-doped part of the phase diagram [22]. The pseudogap signal, which is characterized by a fast decay component, persists below $T_C$, indicating the coexistence of superconducting and pseudogap components. In addition it was shown that the QP dynamics of the pseudogap response is different from that of the superconducting gap. Specifically, for the superconducting component, the decay rate is strongly dependent on the excitation power especially in the weak perturbation regime, while the pseudogap response is nearly independent of the power. As another approach, in a work done by Liu *et al.* [56], the responses from the pseudogap and superconducting phase have been differentiated through polarization- and wavelength - dependent measurements.

While the weak perturbation was thought to be important for clarifying different responses of the pseudogap and superconducting phases, in our work it was shown that the phase separation can also be distinguished under high power excitations up to the level of ~ 100 μJ/cm$^2$ [62]. Figure 2 presents a comparative study of the pump-probe data from the YBCO film at 50 K and 110 K under different power excitations. As seen in Figs. 2(a) and 2(b), the power-dependences of QP dynamics in the superconducting phase and pseudogap phase are different. Under relatively weak excitation, the QP relaxation in the superconducting phase is noticeably slower than in the pseudogap phase [Fig.2 (d)]. With increasing excitation power, the bolometric response becomes more distinct. By properly subtracting the bolometric response [62], the signals remarkably fall into three distinct regimes [Fig. 3(a)]: (1) for $T > T^*$, the bolometric response strongly dominates the transient optical response, (2) below $T_C$, these data show a very rapid positive reflectivity change followed by a slow exponential recovery, and (3) in the intermediate temperature regime of $T_C < T < T^*$, the reflectivity change is progressively weakened. The temperature dependence of such a signal can also be well interpreted with the sum of responses from the superconducting and pseudogap phases [Fig. 3 (b)].

The coexistence of the pseudogap and superconductivity components at temperature below $T_C$ has been observed in several pump-probe experiments, but the relationship between pseudogap and superconductivity has been an outstanding issue for a long time. In a two-color pump-probe study on Bi$_2$Sr$_2$CaCu$_2$O$_{8+\delta}$, Toda *et al.* identified a correlation between superconducting and pseudogap QPs [70]. In the recent study on Nd$_{2-x}$Ce$_x$CuO$_{4+\delta}$, it has been argued that there is an emergent signature of competing order above $T_C$ [38]; under weak perturbation the $\Delta R/R$ was shown to change sign at $T_C$. This negative signal persists up to the temperature $T^*$. The detailed analysis of the coupling between the superconductivity and pseudogap components indicates a repulsive interaction between the two orders. Broadband pump-probe spectroscopy has also been carried out by Coslovich *et al.* to systematically study this issue on the sample of $Bi_2Sr_2Ca_{0.92}Y_{0.08}Cu_2O_{8+\delta}$ [40]. Employing the algorithm of singular value decomposition (SVD), they managed to isolate the spectrotemporal fingerprints of pseudogap and superconductivity components from the broadband pump-probe data. It was concluded that the temporal function for the pseudogap phase is temperature dependent. By analyzing the correlation between the superconducting and pseudogap components within the time-dependent Ginzburg-Laudau model they also argued for a weak competing relation between the superconducting phase and the pseudogap phase.

To elucidate the correlation between pseudogap and superconducting components we investigated the nearly optimally-doped YBCO films by employing broadband pump-probe spectroscopy. As seen in Fig.4, the transient trace is strongly dependent on the probe wavelength. The traces consist of the responses from the superconducting phase, pseudogap, and a bolometric background under strong excitation. We were able to disentangle contributions from these components with the method of SVD. Figures 4 (b)-(e) depict the spectrotemporal signatures of the superconducting and pseudogap components. As can be seen, the temporal evaluations of the superconductivity and pseudogap signals suggest a repulsive interaction, which is very similar to that observed in $Bi_2Sr_2Ca_{0.92}Y_{0.08}Cu_2O_{8+\delta}$ [40]. The spectral signatures of these two components show the same sign with the probe photon energy in the range of 1.8~2.4 eV, but opposite signs in other range, as shown in Fig.4(c). This important observation may explain the discrepancy of why the signals of superconducting and pseudogap components can have either the same [20] or different signs [38, 56] in single-color pump-probe spectroscopy. It is interesting to speculate whether the peak feature appearing around 2.0 eV could be related to the coupling between QPs and charge-transfer excitations.

Besides the recovery dynamics, the pseudogap phase has also been characterized by studying anomalies in coherent phonon dynamics [57]. The broadband spectral bandwidth of ultrashort pulses employed in pump-probe experiments enables the excitation of coherent phonon states. They are manifested in pump-probe traces as a modulation of the optical dielectric function at the frequency corresponding to lattice phonons [41, 63, 88, 89]. In YBCO, coherent optical phonons have been observed decades ago [63]. The oscillation frequencies, as estimated from Fourier transformation, are in good agreement with the Raman spectroscopic data, and the temperature-dependent Ba and Cu phonon modes have been clearly identified as well [89]. In the normal state, two crossover temperatures of those modes have been attributed to the inhomogeneity of pseudogap phase [57]. Different mechanisms have been proposed to explain the coherent phonons in cuprates [90].

The charge density wave (CDW) in cuprates is yet another collective mode that has attracted great attention for ultrafast optical studies of cuprates recently. The dynamics and photo-induced melting of the CDW phase have been investigated in other systems by different time-resolved methods [91-96]. In cuprates, the challenge is to understand if the CDW competes with superconductivity. Following this goal, recently, ultrafast spectroscopy has been employed to investigate the dynamic CDWs in the under-doped cuprates [34, 37]. The pump-probe and transient grating spectroscopies are used to disentangle the amplitude and phase of the reflectivity change induced by optical excitation, respectively. Gedik *et al.* have reported the presence of an oscillation in the pump-probe traces from $La_{1.9}Sr_{0.1}CuO_4$ samples [34]. The amplitude of the oscillation was assigned to the amplitude modes of the CDWs since their temperature and doping dependences are very similar. In a recent work, Hinton *et al.* conducted a similar study in the under-doped YBCO sample and reported the observation of coherent oscillations associated with the CDW order [37]. The oscillation amplitude was enhanced upon cooling below the superconducting critical temperature. Despite the fact that the amplitude of oscillation shows a temperature dependence similar to CDWs, the parameters of the oscillation component lack certain features of an amplitude mode. In particular, the frequency shift with temperature predicted for CDWs is absent. This new mode has been assigned to the acoustic phonons whose wave vector matches that of the CDWs. Considering the similarity of the collective modes observed in YBCO and LSCO, Hinton *et al.* suggested that the oscillation observed in LSCO [34] may not be an amplitude mode of the CDWs but rather a result of mixing-down of acoustic modes to zero wave vector [37].

### 3.3. Photo-induced phase transition

In most pump-probe experiments designed to study the electronic responses of different phases in cuprates, the excitation power has been intentionally kept low to be in a weak excitation regime to minimize laser-induced extrinsic changes to the samples. However, upon increasing excitation power beyond a certain level, the photo-thermal effect becomes important. In general, the laser pulses may induce both steady and transient thermal effects. On one hand, the steady thermal effect may cause a baseline change in the transient optical traces, which can be avoided by decreasing the repetition rate of the excitation laser. On the other hand, the non-thermal effect has attracted more attention since it relates to the electronic properties in cuprates. Naturally, both thermal and non-thermal effects induced by laser excitation may cause phase transitions in cuprate superconductors [23, 29, 55, 58].

Kusar *et al.* performed a study on the laser-driven vaporization of the superconducting condensate in LSCO samples [58]. Dramatic divergence in QP dynamics was observed when the excitation power surpasses a threshold value of about 2000 photons per Cu. They observed that the dynamics of photo-excited QPs at 4.5 K is similar to that at the normal state, indicating melting of the superconducting condensates. It was suggested that the temperature dependence of $\Delta R/R$ can be modeled according to the Mattis-Bardeen formula [58, 97]

$$\Delta R/R \propto \frac{2\Delta(\mathrm{T})}{\hbar\omega}\ln(\frac{1.47\hbar\omega}{\Delta(\mathrm{T})}), (2)$$

with the temperature-dependent superconducting gap $\Delta(T)$ and excitation photon energy $\hbar\omega$. The melting dynamics has also been analyzed with respect to the issue of pairing glue. Kusar *et al.* argued that the QP dynamics, especially the sub-ps melting process of superconducting condensates, is likely to be related to pair formation by phononic glues [58]. The laser-induced destruction of superconductivity has been studied in several other cuprates as well [55]. Evidence of a non-thermal phase transition from the superconducting to normal state has been reported, which proposed that the first-order photo-induced phase transition is driven by the shift of the chemical potential [23]. The melting processes of superconductivity and CDWs in cuprates have been comparatively investigated by a technique with optical pump and terahertz probe [94]. Due to the relatively large magnitude of a CDW gap, the melting process is largely governed by the electronic processes.

Motivated by the above considerations, we conjecture that it is likely for the photo-induced phase transition to occur in our YBCO films as well. For this purpose we have analyzed the decay traces in Fig. 2(a) with a bi-exponential function $\Delta R/R(t) = A_1 e^{-t/\tau_1} + A_2 e^{-t/\tau_2} + A_0$ and focused on the fast component caused by the electronic response [Fig. 2(c)]. The resulting fluence dependences of the amplitude and lifetime parameters are plotted in Fig. 2(d). As can be seen that a discontinuity in the fluence dependence is clearly observed; above the threshold value, the fluence dependence of the signal amplitude departs from the linear trend [23, 55]. In a similar excitation regime, the lifetime becomes insensitive to the excitation power. Such behaviors have been regarded as evidences for the photo-induced transition from the superconductivity to the normal state [23, 55]. Under high-power excitation, however, the decay dynamics of QPs at 50 K follows a similar trace as that at 110 K [Fig. 2(f)], indicating the melting regime for superconductivity.

Besides suppressing superconductivity by the NIR pulses at 800 nm, the group led by A. Cavalleri has performed a series of experiments to enhance the superconductivity by mid-IR pulses [29, 71, 98]. It is well known that superconductivity in cuprates coexists with different competing orders; therefore suppressing one order by an external mean may result in enhancing another one. Specifically, Cavalleri *et al.* used THz pulses to induce superconductivity in under-doped and originally non-superconducting $La_{1.8-x}Eu_{0.2}Sr_xCuO_4$ samples. Upon resonantly exciting an in-plane Cu-O stretch mode with mid-IR pulses, the signature of superconductivity, i.e. a Josephson plasma resonance, appears in the THz conductivity spectrum. Based on the presented result, it seems that the superconducting state could have been generated within 1-2 ps, thus opening a new venue for manipulating superconductivity with external photo-excitation. Additionally, such a scheme has been tested in under-doped YBCO crystals. Again, transient inhomogeneous superconductivity was possibly obtained at the temperature far above the critical temperature for steady superconductivity; the authors claimed to have reached the highest temperature for inhomogeneous superconductivity in $YBa_2Cu_3O_{6.45}$, which is above room temperature [71]. Based on the experiments with broader THz spectral coverage Cavalleri *et al.* suggested that the enhancement of superconductivity in YBCO might have been achieved by redistributing the interlayer coupling. These intriguing results will certainly rejuvenate interest in this field and may provide unprecedented information about the high-temperature superconductivity in cuprates.

## 4. Perspectives

In the past three decades, technical improvements in the technology of ultra-short pulse lasers and the scientific advances in high temperature superconductivity have been rapidly developing. The promising potentials for their marriage have been clearly witnessed in the past few years. Besides cuprate superconductors, a variety of interesting superconductors including BCS superconductors [59, 66], superconductivity in heavy Fermion compounds [19, 65, 99, 100], and iron-based superconductors [35, 36, 39, 68, 86, 101] have also been subjected to ultrafast pump-probe studies. In this paper, we have reviewed the results on non-equilibrium phenomena in cuprate superconductors obtained by optical pump-probe measurements. The changes in transient reflectivity are extremely sensitive probes of QP dynamics at the Fermi level, providing a wealth of information on the pairing glues and competing orders which are the keys to understand high-temperature superconductivity. On the other hand, the transient reflectivity seems to be rather insensitive to magnetic structure which is essential to understand the electron correlations in high temperature superconductivity. The rapidly developing techniques, such as time-domain spin- and angle-resolved photoemission spectroscopy, X-ray absorption, circular dichroism, and diffraction experiments, can likely contribute to the better understanding of the magnetism in such materials. The extension of wavelength coverage from X-rays to THz will enable selective excitations of well-defined modes [33, 102]. Undoubtedly, all these improvements will be vital for finally uncovering the microscopic mechanism of high-temperature superconductivity.

**Final note and Acknowledgement**

By no means does this paper present a comprehensive review of high Tc superconductivity with ultrafast optical spectroscopies. We, and as such the authors, have not attempted to review the enormous body of published results. Instead, the authors have focused on a few selected results to illustrate the concepts, methodologies and physics behind described phenomena. We apologize for possible omissions.

The work at Nanjing University is supported by the National Basic Research Program of China (2013CB932903 and 2012CB921801, MOST), the National Science Foundation of China (91233103, 61108001, 11227406 and 11321063). The authors would like to acknowledge to the members in the group for their cooperation, discussion and useful comments: B. Gary, Wei Guo, Shenghua Liu, Bin He, and Ye Wang. C.Z. acknowledges Professor Jianxin Li for stimulated discussions and D. Meyers for reading the manuscript. JC was supported by DOD-ARO Grant No. 0402-17291.

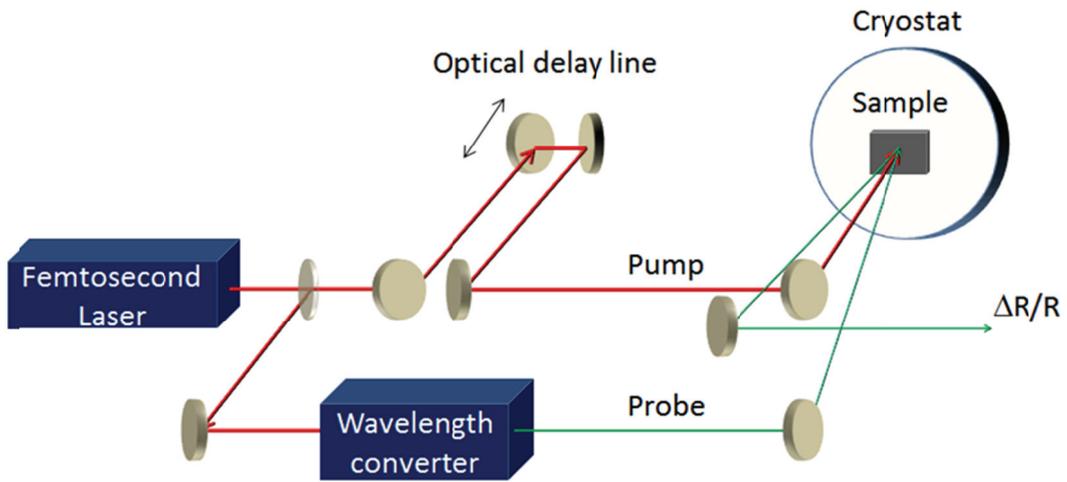

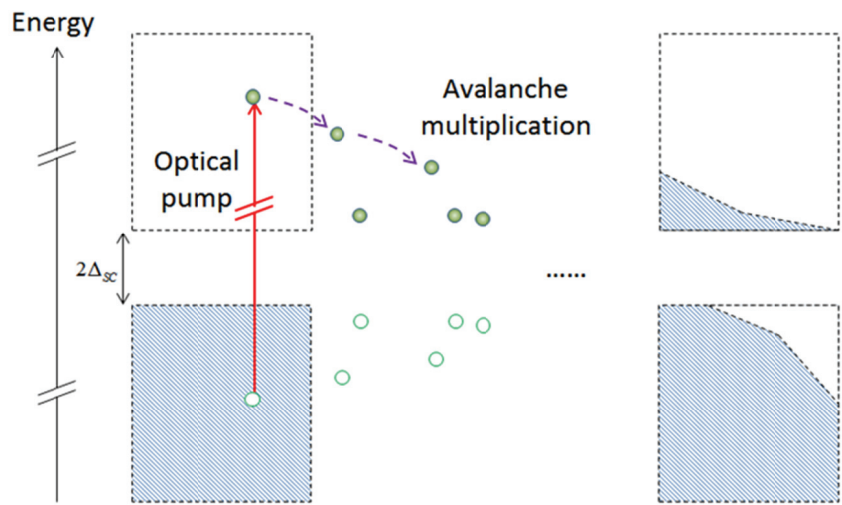

Figure 1, (a) A schematic diagram of the pump-probe spectroscopy. (b) A schematic illustration of the ultrafast photo-excitation processes in cuprates. One absorbed photon generates a QP with high energy which loses its energy by creating multiple QPs through the process of avalanche multiplication.

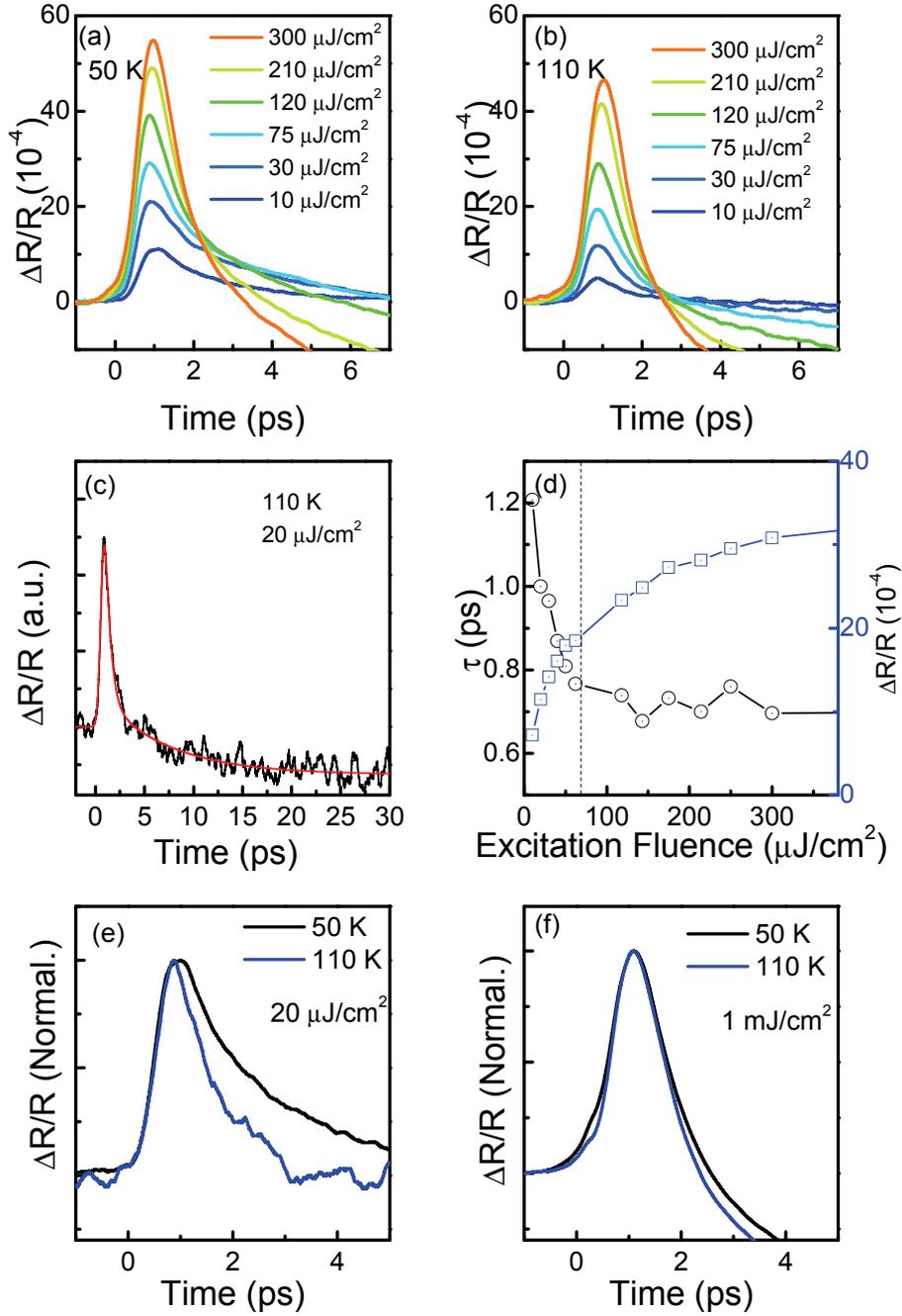

Figure 2, Pump-probe traces recorded at 50 K (a) and 110 K (b) under different excitation flux. (c) An example of the fitting process on a trace recorded at 110 K under excitation fluence of 20 μJ/cm$^2$. (d) The amplitude and fast decay lifetime recorded at 50 K are plotted as a function of excitation fluence. The dashed line is an eye guide for the critical excitation fluence, above which the signal amplitude departs from the linear dependence on the excitation flux. (e) & (f) compare the normalized pump-probe traces recorded at 50 K and 110 K under excitation fluences of 20 μJ/cm$^2$ and 1 mJ/cm$^2$, respectively. The experiments were carried out with the pump at 400 nm and probe at 800 nm.

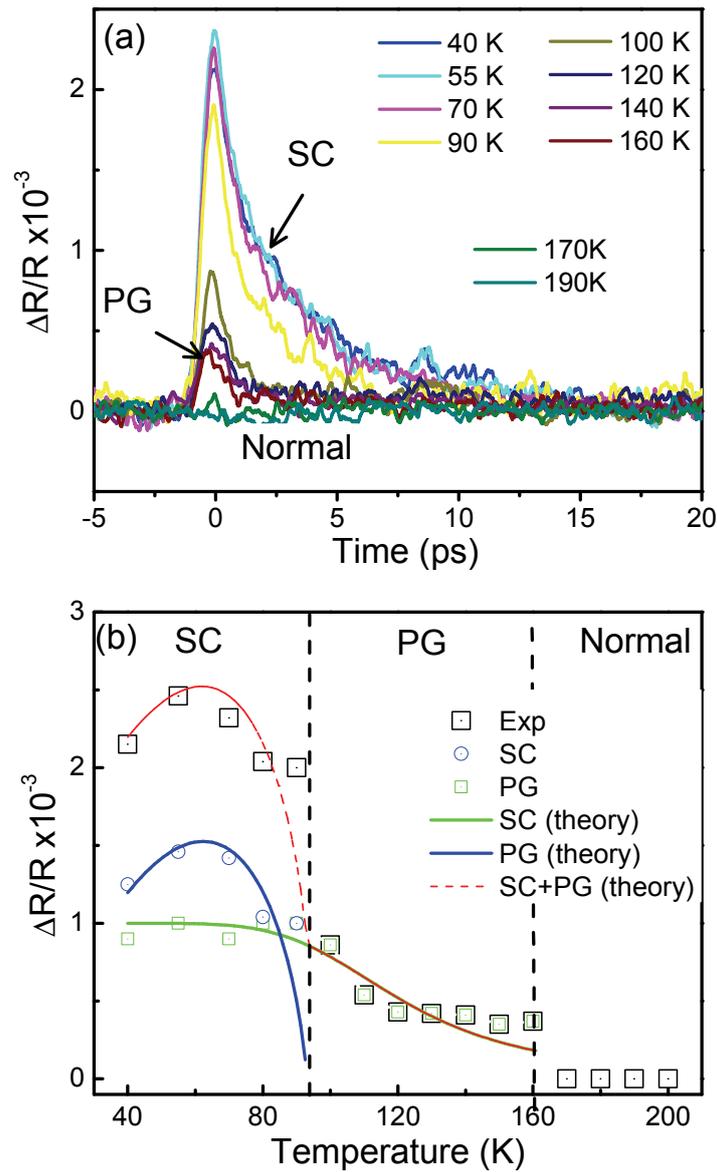

Figure 3, (a) The rescaled pump-probe traces obtained by subtracting the bolometric response recorded from the YBCO film at different temperatures. (b) The temperature-dependent amplitudes of the fast and slow components are compared with the theoretical modeling with a temperature-independent pseudogap and a temperature-dependent superconducting gap. The blue and green solid lines represent the curves fitting to the superconducting gap [25, 58] and pseudogap [25, 68], respectively, while the red dashed line is their summation. The dashed lines indicate the transition temperatures for superconductivity and pseudogap derived from the transport data. The experiments were carried out with the pump at 400 nm and probe at 800 nm [62].

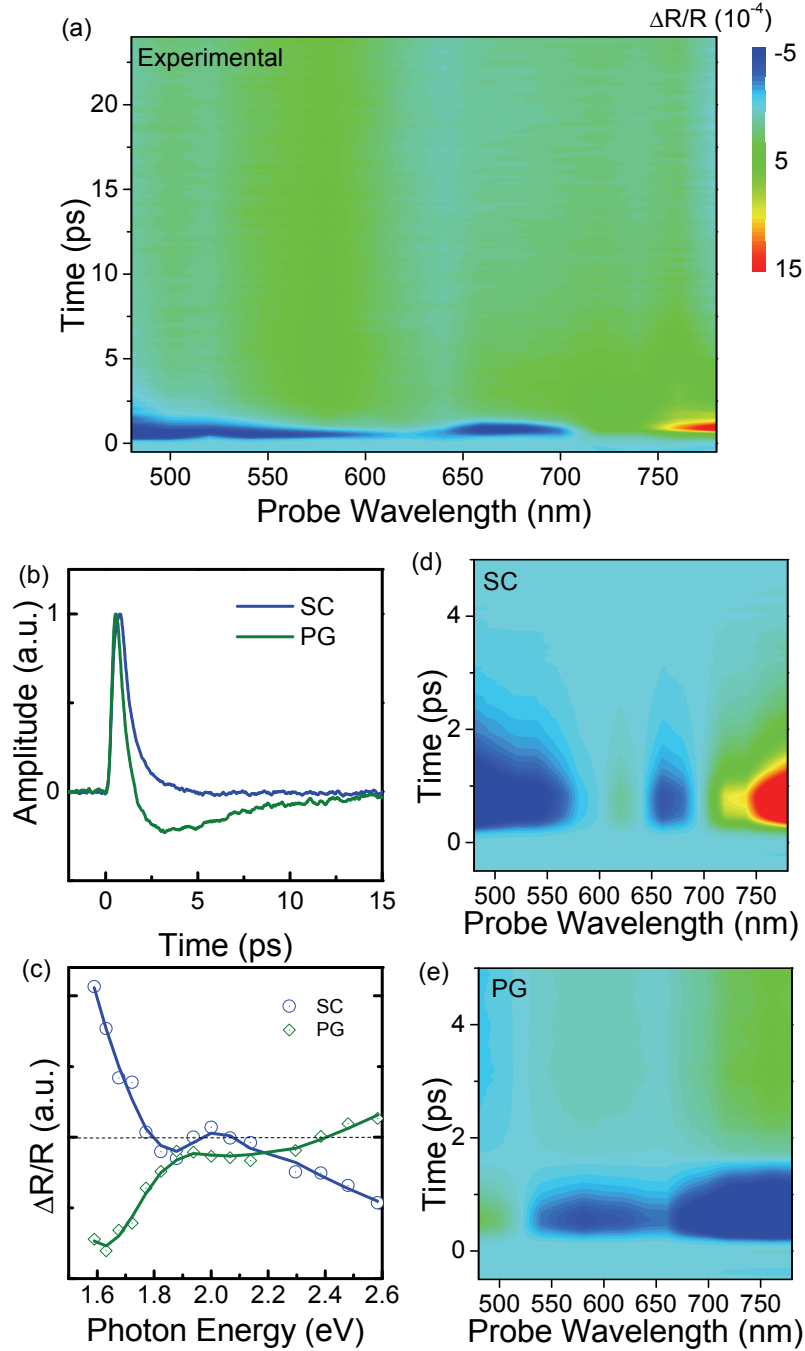

Figure 4. Broadband pump-probe spectroscopic studies on the YBCO films. (a) The values of $\Delta R/R$ are plotted as functions of delay time and probe wavelength. Experimental data recorded at 30 K with the pump at 800 nm with an excitation fluence of ~ 25 uJ/cm$^2$. (b) and (c) plot temporal and spectral functions of the superconducting and pseudogap components reconstructed with the SVD. (d) and (e) The time-energy matrix of the superconducting and pseudogap components plotted in the same scale as in (a).